\documentclass{article}
\usepackage{LaThuileFPSpro}
\begin{document}
\title{ 
LEP1 HEAVY FLAVOUR ELECTROWEAK PHYSICS
AND 2-FERMION PROCESSES AT LEP2
}
\author{
Thomas Sch\"orner-Sadenius\\
{\em CERN, Division EP, 1211 Geneva 23, Switzerland} and\\
{\em Hamburg University, Luruper Chaussee 149, 22761 Hamburg, Germany}
  }
\maketitle

\baselineskip=11.6pt

\begin{abstract}
Results from heavy flavour electroweak physics at LEP1 are reviewed together
with measurements of 2-fermion processes at LEP2. For the former measurements
the emphasis is on analyses of heavy quark forward-backward 
asymmetries, $A_{{\mathrm FB}}^{{\mathrm b,c}}$, and on 
the partial decay widths of the ${\mathrm Z^0}$ to heavy 
quarks, $R_{{\mathrm b,c}}$. The measurements of
the heavy quark asymmetries
are used to extract the effective electroweak mixing angle 
for leptons, $\sin^2\theta_{{\mathrm eff}}^{{\mathrm lept}}$. 
A 2.9$\sigma$ discrepancy between measurements of this quantity from
${\mathcal A}_{\mathrm l}({\mathrm SLD})$ and from 
$A_{{\mathrm FB}}^{\mathrm b}$ is observed.
The 2-fermion processes at LEP2 are used to place stringent limits on physics 
processes beyond the Standard Model. However, all 
measured quantities are in reasonable
agreement with the Standard Model expectations, and all 
calculated limits are well 
above the highest LEP2 center-of-mass energies.\\
\end{abstract}
\newpage
\section{Overview}
Although the LEP experiments finished data taking in 2000, the analysis efforts are still strong 
in all four collaborations. In this article I will give an overview of the status of 
two rather distinct experimental areas, namely the heavy flavour electroweak measurements at LEP1 
and 2-fermion processes at LEP2. Whereas the former class of analyses basically serve as a test of
the electroweak Standard Model, with the main interest on the extraction of the weak mixing angle
$\sin^2\theta_{{\mathrm eff}}$, the LEP2 2-fermion measurements can be used for generating limits on 
models for new physics beyond the Standard Model. 

This article is structured as follows: Section~\ref{hf} is devoted to the LEP1 heavy flavour
measurements. After a short overview of general issues and experimental techniques
the results of the partial decay width and forward-backward asymmetry measurements 
for heavy quarks are shown. At the end of the section, an interpretation of these results 
in terms of the Standard Model, namely the electroweak 
mixing angle $\sin^2\theta_{{\mathrm eff}}$, is given.
Section~\ref{2f} reviews measurements of 2-fermion processes at LEP2. A large part of the section
is devoted to the interpretation of the measurements in terms of new physics models, such as
contact interactions, ${\mathrm Z'}$ bosons or low scale gravity in large extra dimensions. 
Section~\ref{conclusion} concludes the paper and gives an outlook.

\section{\label{hf}LEP1 Heavy Flavour Electroweak Physics}
\subsection{Introduction}
Measurements of ${\mathrm b}$ and ${\mathrm c}$ quark final 
states at LEP allow detailed insights into the properties of
the electroweak Standard Model. The measurements considered here 
are a) the heavy flavour partial decay
widths of the ${\mathrm Z^0}$, $R_{\mathrm b}$ and $R_{\mathrm c}$, and b) the 
heavy quark forward-backward asymmetries, $A_{{\mathrm FB}}^{\mathrm b}$
and $A_{{\mathrm FB}}^{\mathrm c}$.

$R_{\mathrm q}$ is defined as the decay width of the ${\mathrm Z^0}$ to quarks of 
flavour ${\mathrm q}$, normalized to the total 
hadronic decay width, and is proportional to the sum of the squared vector and axial vector 
couplings of the ${\mathrm Z^0}$:
\begin{equation}
R_{\mathrm q} := \frac{\Gamma^{\mathrm q}}{\Gamma^{had}} \propto g_{V{\mathrm q}}^2 + g_{A{\mathrm q}}^2,
\end{equation}
$R_{\mathrm q}$ is therefore sensitive to vertex corrections 
to the ${\mathrm q}\overline{{\mathrm q}}{\mathrm Z^0}$ 
vertex and might thus serve to
detect signs of new physics modifying these vertices. 
Conceptually the measurement of $R_{{\mathrm b,c}}$ is simple: 
One has to count the number of events with 
${\mathrm b}\overline{{\mathrm b}}$ or ${\mathrm c}\overline{{\mathrm c}}$ final states 
and normalize to the number of all hadronic final 
state events. So the largest challenge, besides understanding the rather 
involved systematic uncertainties, is
providing an efficient, clean and well-understood ${\mathrm b}$ or ${\mathrm c}$ flavour tag. 

The forward-backward asymmetry measurements are more complicated. 
The asymmetry for flavour ${\mathrm q}$ is defined as
\begin{equation}
A_{{\mathrm FB}}^{\mathrm q} := \frac{N^{\mathrm q}_{\mathrm F}-N^{\mathrm q}_{\mathrm B}}{N^{\mathrm q}_{\mathrm F}+N^{\mathrm q}_{\mathrm B}}
\end{equation} 
where $N^{\mathrm q}_{\mathrm F}$ ($N^{\mathrm q}_{\mathrm B}$) 
is the number of events of flavour ${\mathrm q}$ in which the initial quark goes into the 
forward (backward) hemisphere. The forward hemisphere is defined by the electron beam direction. 
Since at the ${\mathrm Z^0}$ pole
\begin{equation}
A_{{\mathrm FB}}^{{\mathrm q,0}} = \frac{3}{4} {\mathcal A}_{\mathrm e} {\mathcal A}_{\mathrm b}
\end{equation}
with
\begin{equation}
{\mathcal A}_{\mathrm f} = 2 \frac{g_{V{\mathrm f}} \cdot g_{A{\mathrm f}}}{g_{V{\mathrm f}}^2 + g_{A{\mathrm f}}^2}
\end{equation}
the asymmetries provide direct access to the electroweak mixing angle $\sin^2\theta_{{\mathrm eff}}$
via the relation $g_V/g_A = 1 - 4 Q \cdot \sin^2\theta_{{\mathrm eff}}$. Due to the charge factor $Q$ in the 
former relation, the sensitivity of the asymmetries to the 
heavy quark couplings ${\mathcal A}_{\mathrm q}$ is much reduced with respect 
to the electron coupling ${\mathcal A}_{\mathrm e}$ (even more for 
${\mathrm b}$ than for ${\mathrm c}$ quarks). 
So the measurements serve in fact as a 
determination of the mixing angle for electrons (or leptons, assuming lepton universality), 
$\sin^2\theta_{{\mathrm eff}}^{{\mathrm lept}}$.

The SLD ${\mathrm e^+e^-}$ collider, in contrast to LEP, has the possibility to 
polarize the electron beam so that 
additional measurements become available. One can for example, 
from a measurement of the left-right
forward-backward asymmetry, $A_{{\mathrm LRFB}}$, directly extract the heavy 
flavour couplings ${\mathcal A}_{{\mathrm b,c}}$ which are not accessible at LEP.
SLD measurements are also used in the heavy flavour fits 
that will be covered in the next sections. 

The measurements of the partial decay widths\cite{rblit} are final. 
The latest contributions to the forward-backward asymmetry measurements
were presented at the 2002 summer conferences: OPAL prepared a new 
measurement of the ${\mathrm b}$ quark
asymmetry using an inclusive charge tag\cite{opalafbb} which superseeded their old 
result and substantially reduced the statistical and systematic uncertainties. 
DELPHI presented new preliminary measurements of the 
lepton-tagged ${\mathrm b}$ and ${\mathrm c}$ asymmetries\cite{delphiafbl} 
which included new systematic studies.
It is therefore fair to say that the LEP1 heavy flavour electroweak measurements are slowly coming
to an end. The only missing pieces are a lepton-tag asymmetry measurement from OPAL and the
inclusive ${\mathrm b}$ asymmetry measurement from DELPHI. In addition, 
some SLD measurements are also only 
preliminary ($A_{{\mathrm LRFB}}$, $R_{\mathrm c}$).

\begin{figure}[t]
 \vspace{9.0cm}
\includegraphics{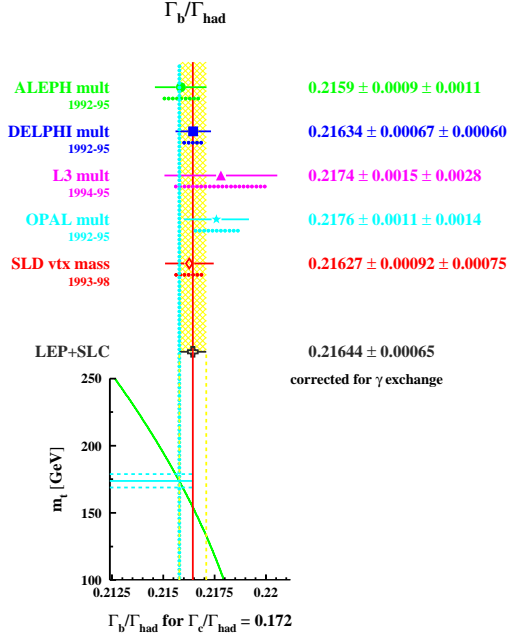}
 \caption{\it A compilation of measurements of $R_{\mathrm b}$ from LEP and SLD.
Also the Standard Model expectation, which depends on 
the mass of the top quark $m_{\mathrm t}$, is shown.
\label{rb} }
\end{figure}

\subsection{\label{hftools}Tools: Flavour and Charge Tags}
As became apparent in the previous section, the heavy flavour measurements rely on flavour tagging, 
and, in case of the asymmetry measurements, also on tagging the charges of the outgoing quarks. The 
determination of these quantities is facilitated by the usually 
clear 2-jet structure of hadronic events 
at or around the ${\mathrm Z^0}$ which allows the events to be divided in two hemispheres along the 
plane perpendicular to the thrust axis. The flavour and charge tagging tools can then 
be applied in the two hemispheres independently. 
Due to this basic structure nearly all information needed for the extraction of the partial decay
widths and asymmetry values can be taken from the data 
without relying too much on Monte Carlo input. 

Heavy flavour events can be tagged using heavy quark secondary decay vertices or high $p_T$ leptons 
from semileptonic decays ${\mathrm b,c} \rightarrow {\mathrm l}$. The variables describing the decay vertices 
or the decay leptons are usually 
combined using likelihoods or artificial neural networks to result in one flavour tagging variable
with high separation power between ${\mathrm b}$ or ${\mathrm c}$ quark hemispheres and the light flavour background, 
respectively. Purities of 
95~$\%$ with efficiencies of 20 to 30~$\%$ can be reached. Another possibility is to 
tag ${\mathrm c}$ hemispheres using ${\mathrm D}^*$ mesons. 

\begin{figure}[t]
 \vspace{7.0cm}
\includegraphics{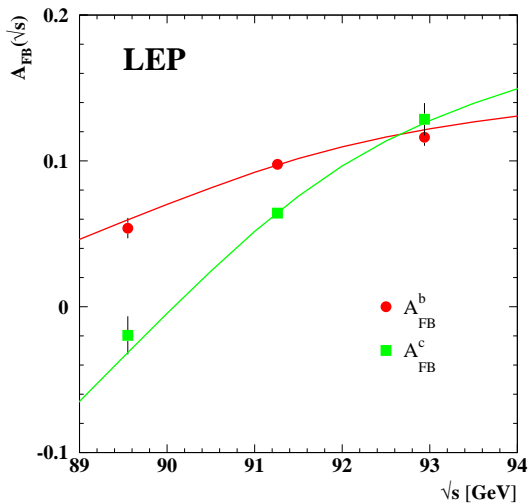}
 \caption{\it The ${\mathrm b}$ and ${\mathrm c}$ quark asymmetries as a function of 
the center-of-mass energy, compared to the ZFITTER Standard Model prediction.
    \label{afbene} }
\end{figure}

The charge of hemispheres which is needed for the forward-backward 
asymmetry analyses can be determined 
from a variety of observables. The OPAL analysis mentioned above uses the jet charge of the 
highest energy jet in the hemisphere, the weighted charge sum of all tracks connected to the 
secondary vertex in the hemisphere and, in addition, the charge of kaons originating from 
cascade decays ${\mathrm b} \rightarrow {\mathrm c} \rightarrow {\mathrm s}$ 
measured using the d$E$/d$x$ information from the 
OPAL central jet chamber. These variables are again combined in an artificial neural network.
Charge tags like the 
one just described are self-calibrating in the sense that their efficiency and purity can be
determined from the data themselves, thus reducing the dependence on external inputs from Monte
Carlo simulations. It turns out that the correlation between the charge 
tags in the two event hemispheres is one of the dominant sources of systematic uncertainty.

In case of lepton-tagged asymmetry measurements, basically the same variables as for the flavour 
tag can be used: lepton momentum and transverse momentum, vertex fit probability, to mention a few, 
together with some charge information from the hemisphere opposite to the one with 
the identified lepton. 

\subsection{\label{hfresults}Results on Heavy Quark Decay Widths and Asymmetries}
In order to get a more precise picture of heavy flavour electroweak physics, the various measurements
of the partial decay widths and asymmetries are combined in a $\chi^2$ minimisation procedure. 
In this fit, the correlations between various measurements due to mutual dependencies and to
common external inputs are taken into account. Together with 
the LEP measurements of $R_{{\mathrm b,c}}$
and $A_{{\mathrm FB}}^{{\mathrm b,c}}$ the SLD measurements of ${\mathcal A}_{\mathrm b}$ 
and ${\mathcal A}_{\mathrm e}$ are used in this fit. 
Some additional auxiliary parameters (charm hadron production fractions, ${\mathrm b}$ 
semileptonic branching ratios) are also determined. 

\begin{figure}[t]
 \vspace{8.0cm}
\includegraphics{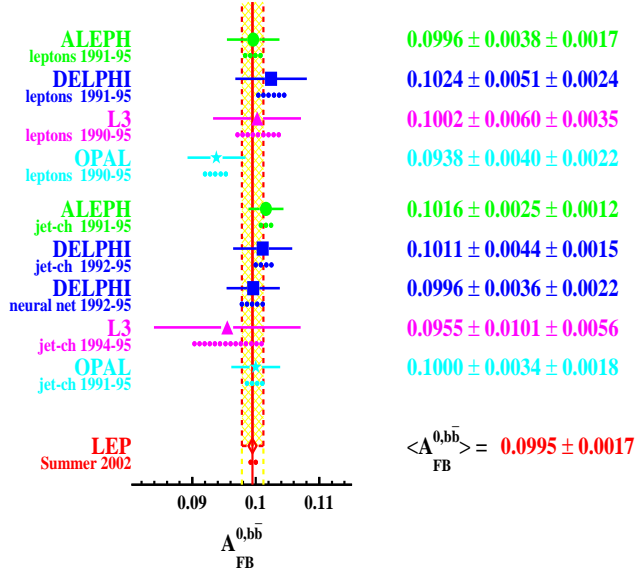}
 \caption{\it A compilation of  
$A_{{\mathrm FB}}^{\mathrm b}$ measurements from LEP and SLD.
\label{afb} }
\end{figure}

Figure~\ref{rb} shows a compilation of measurements 
of $R_{\mathrm b}$ together with the $R_{\mathrm b}$ result of the heavy flavour fit\cite{rblit}. 
The measurements of the four LEP experiments and of SLD are 
in good overall agreement. The fit results in a value $R_{\mathrm b} = 0.21644 \pm 0.00065$.
This value,
indicated by the solid vertical line, is well compatible with 
the Standard Model prediction based on the 
best electroweak knowledge and the Tevatron top mass measurement (left vertical line). 
The various measurements of $R_{\mathrm c}$ are also averaged and result in a 
value of $R_{\mathrm c} = 0.1717 \pm 0.0031$, again in good agreement with the 
theoretical expectation. 

Turning to the asymmetry measurements, first all asymmetry measurements are corrected to three 
distinct energies (the ${\mathrm Z^0}$ pole and $\pm$2~GeV away from it), 
and the averaged asymmetry is 
extracted for these three energies separately. The result is shown in figure~\ref{afbene}. 
Since the variation of the asymmetry with the center-of-mass energy is consistent 
with the Standard Model expectation, all asymmetry measurements are corrected to the ${\mathrm Z^0}$ 
pole, corrected for photon radiation and quark mass effects, and combined. The contributing 
$A_{{\mathrm FB}}^{\mathrm b}$ measurements\cite{opalafbb,delphiafbl,afblit,afblit2} 
and the derived global value of 
$A_{{\mathrm FB}}^{{\mathrm b,0}} = 0.0995 \pm 0.0015 \pm 0.0007$ are shown in 
figure~\ref{afb}. The average
is dominated by the inclusive jet charge analyses from ALEPH, DELPHI and OPAL. For the ${\mathrm c}$ 
asymmetry\cite{delphiafbl,afblit2,afbc}
we find $A_{{\mathrm FB}}^{{\mathrm c,0}} = 0.0713 \pm 0.0031 \pm 0.0018$. 
In both cases the first error is statistical, 
and the second systematic. The common systematic 
uncertainty is 0.0004 (0.0009) for the ${\mathrm b}$ (${\mathrm c}$) 
asymmetry, resulting mainly from corrections
due to gluon radiation.

\begin{figure}[t]
 \vspace{9.0cm}
\includegraphics{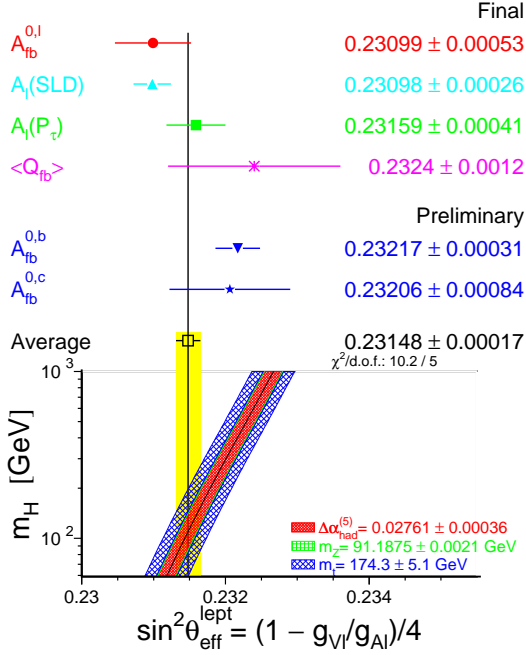}
 \caption{\it A compilation of measurements of the effective electroweak 
mixing angle for leptons, together with the LEP+SLD average.
    \label{sin2theta} }
\end{figure}

\subsection{\label{hffit}Standard Model Interpretation}
One can go one step further using some results of the heavy flavour electroweak fit described
above ($A_{{\mathrm FB}}^{\mathrm b}$ and $A_{{\mathrm FB}}^{\mathrm c}$), 
LEP combined measurements of the 
leptonic forward-backward asymmetry
($A_{{\mathrm FB}}^{\mathrm l}$ and ${\mathcal A}_{\mathrm l}(P_{\tau})$) 
and the SLD ${\mathcal A}_{\mathrm l}$. Expressing 
all these quantities in terms of the 
vector and axial vector couplings of the ${\mathrm Z^0}$, $g_V$ and $g_A$, 
one can extract the effective
electroweak mixing angle for leptons, $\sin^2\theta_{{\mathrm eff}}^{{\mathrm lept}}$. 

As can be seen from figure~\ref{sin2theta}, the resulting single 
values for this quantity fall roughly 
in two classes: the leptonic measurements ($A_{{\mathrm FB}}^{\mathrm l}$, 
${\mathcal A}_{\mathrm l}(P_{\tau})$, ${\mathcal A}_{\mathrm l}({\mathrm SLD})$), which are
dominated by the SLD number, tend to be slightly low, whereas the hadronic or inclusive results, 
in particular for $A_{{\mathrm FB}}^{\mathrm b}$, are rather high. The observed discrepancy 
of 2.9$\sigma$ results in a $\chi^2$ of 10.2/5, corresponding to a 7~$\%$ probability. 

The final $\sin^2\theta_{{\mathrm eff}}^{{\mathrm lept}}$ value is 0.23148$\pm$0.00017, 
where the uncertainty is dominated
by statistics. This combined value of the mixing angle, interpreted in the framework of 
the Standard Model, prefers a Higgs mass slightly above 100~GeV, whilst the 
$A_{{\mathrm FB}}^{\mathrm b}$ measurements alone suggest a heavier Higgs of about 400~GeV.  

\section{\label{2f}2-Fermion Processes at LEP2}
\subsection{Introduction}
The emphasis in measurements of 2-fermion processes at LEP2 is on searching for signs of 
new physics in the interference of new processes with the off-resonance ${\mathrm Z^0}/\gamma$ exchange.
There are, however, a few general issues that have to be considered before deriving general 
results for 2-fermion processes. 
The most prominent problem is initial state radiation. Electron-positron 
annihilation events at center-of-mass energies above the ${\mathrm Z^0}$ pole tend to radiate
an energetic photon in the initial state, forcing the propagator back to the ${\mathrm Z^0}$ mass. 
There is therefore a large contribution to the overall event sample at an effective 
center-of-mass energy corresponding to the ${\mathrm Z^0}$. These events, which do not contain any 
information not known from LEP1 physics, are rejected using a cut on the effective 
center-of-mass energy $\sqrt{s'}$ of the order of 0.85-0.9$\sqrt{s}$. 
In addition to this complication, corrections have to be applied
in order to compensate for different signal definitions and different treatments of ISR-FSR 
interference before the data of different experiments can be combined.

\begin{figure}[t]
 \vspace{7.5cm}
\includegraphics{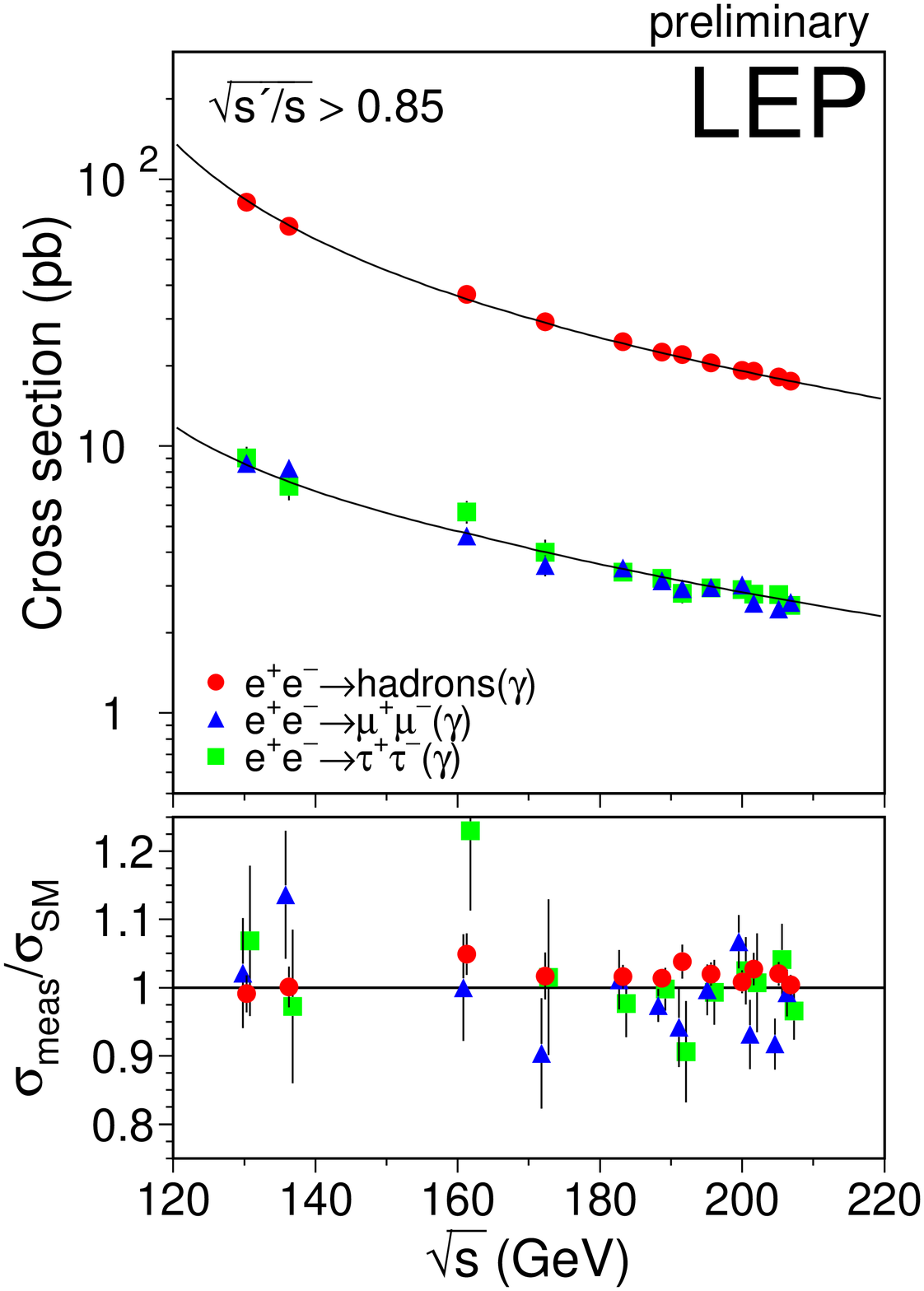}
\includegraphics{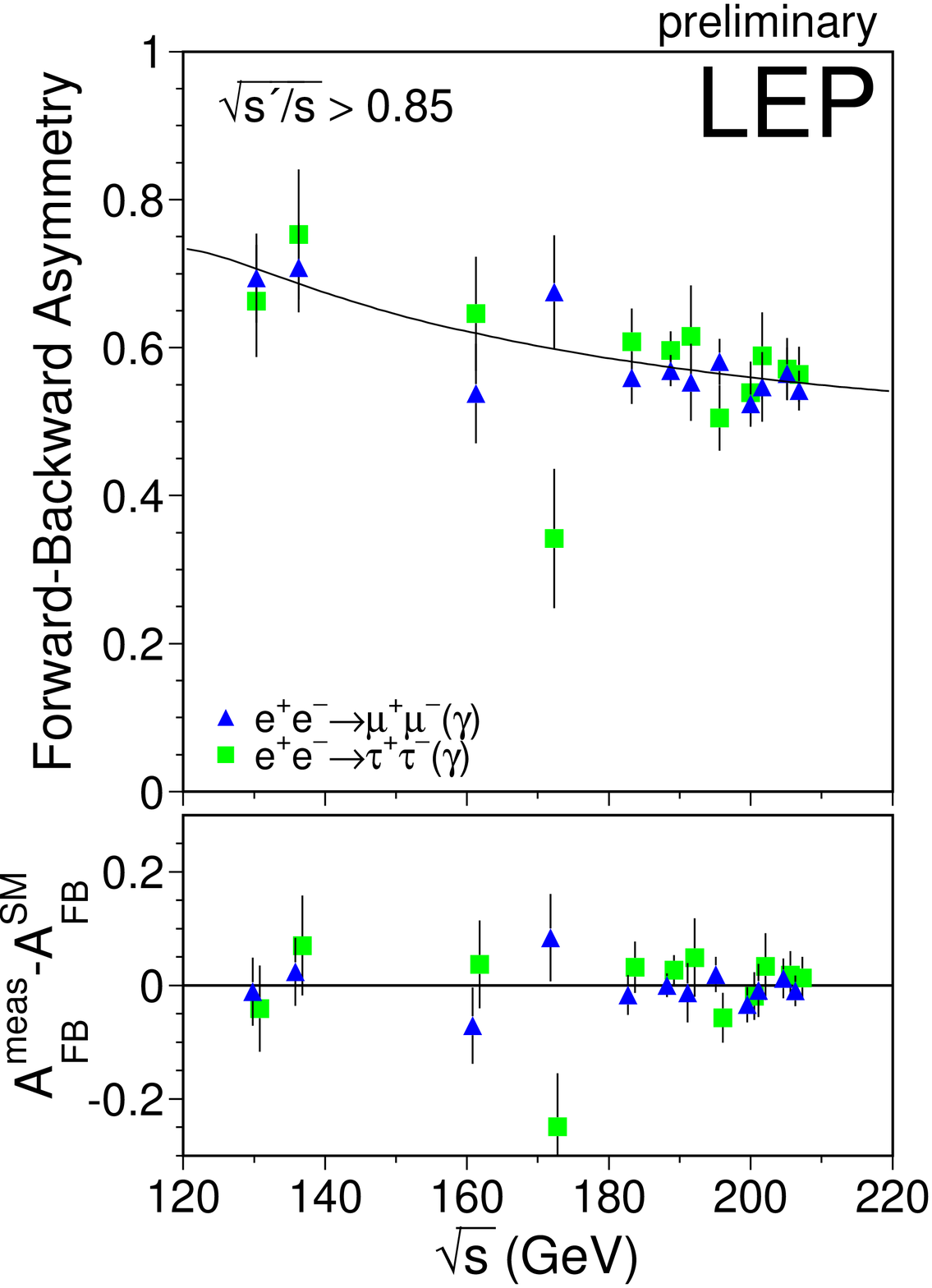}
 \caption{\it The total cross-section and the forward-backward asymmetry for hadronic, 
$\mu$ and $\tau$ final states as a function of the center-of-mass energy, $\sqrt{s}$.
    \label{2fcross} }
\end{figure}

\subsection{\label{2fmeasure}Measurements}
The conceptually simplest measurement of 2-fermion processes is clearly that of 
the total cross-section as a function of the center-of-mass energy $\sqrt{s}$. After the 
above mentioned corrections have been applied, all 
measurements for quark, $\mu$ and $\tau$ final states from the 
four LEP collaborations are combined in a single $\chi^2$ fit to give the average cross-sections
(see\cite{2flit1} and the LEP EW 2f subgroup 
web page\cite{2fweb} for an overview on all measurements used). The results
are shown as a function of $\sqrt{s}$ in the left part of figure~\ref{2fcross}. 
The bottom part of the figure also shows 
the ratio of the data to the Standard Model expectation, indicating an overall good agreement 
for the leptonic channels. The hadronic measurements are slightly low (1.7$\sigma$).
The overall $\chi^2$ is 160/180. 
The same data are also used for the extraction of the forward-backward asymmetry; the result, 
again as a function of $\sqrt{s}$, is shown in the right half of figure~\ref{2fcross}. 
Here, the overall description of all data is satisfactory. 

\begin{figure}[t]
 \vspace{9.0cm}
\includegraphics{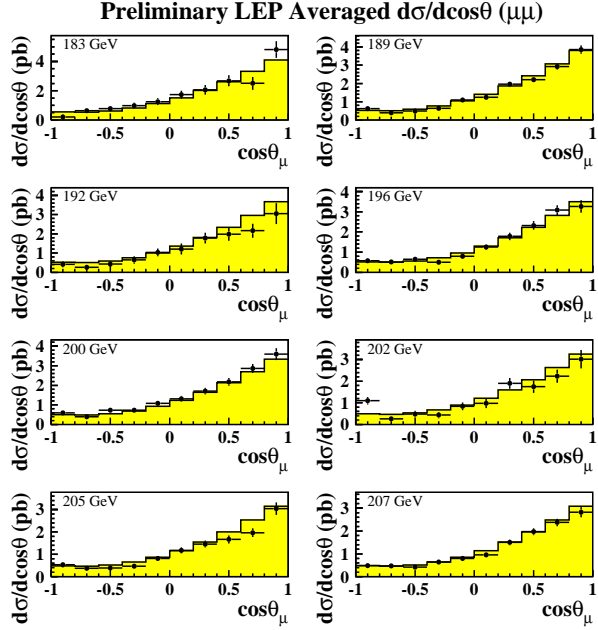}
 \caption{\it Differential cross-section as a function of $\cos \theta$ 
for ${\mathrm e}^+{\mathrm e}^- \rightarrow \mu^+\mu^-$
   for different center-of-mass energies. The data excess in the leftmost bin at 202~GeV is also 
present in the $\tau$ channel. 
    \label{differential} }
\end{figure}

In a next step, differential cross-sections as a function of the scattering angle $\cos\theta$ 
are extracted for all three lepton generations in seven regions of the center-of-mass energy
between 189 and 207~GeV. For the ${\mathrm e^+e^-}$ final states, the additional $t$ channel contribution
which is not present for the other lepton generations leads to a divergence of the cross-section 
towards $\cos\theta = 1$. The description of the data is satisfactory, except for the lowest 
bins for $\mu^+\mu^-$ and $\tau^+\tau^-$ final states at $\sqrt{s} =$~202~GeV. However, 
in these bins there are only low statistics, and for all 
other (lower and higher) energies this feature is not present, see figure~\ref{differential}. 

For heavy flavour quarks, the partial decays widths 
$R_{\mathrm q}$=$\sigma_{{\mathrm q}\overline{{\mathrm q}}}$/$\sigma_{\mathrm hadrons}$ 
and the forward-backward asymmetries are 
measured separately. For $R_{\mathrm c}$, the only measurement is provided by ALEPH. 
The agreement between data and the ZFITTER Standard Model expectation 
is acceptable, except perhaps
for $R_{\mathrm b}$ which tends to be low by about 2$\sigma$ over all energies. 

\subsection{\label{2finter}Interpretation in Non-standard Models}
The data described above are interpreted in several physics models: electron-lepton and 
lepton-quark contact interactions, ${\mathrm Z'}$ bosons, and low scale gravity in large extra 
dimensions. 

\begin{figure}[t]
 \vspace{9.0cm}
\includegraphics{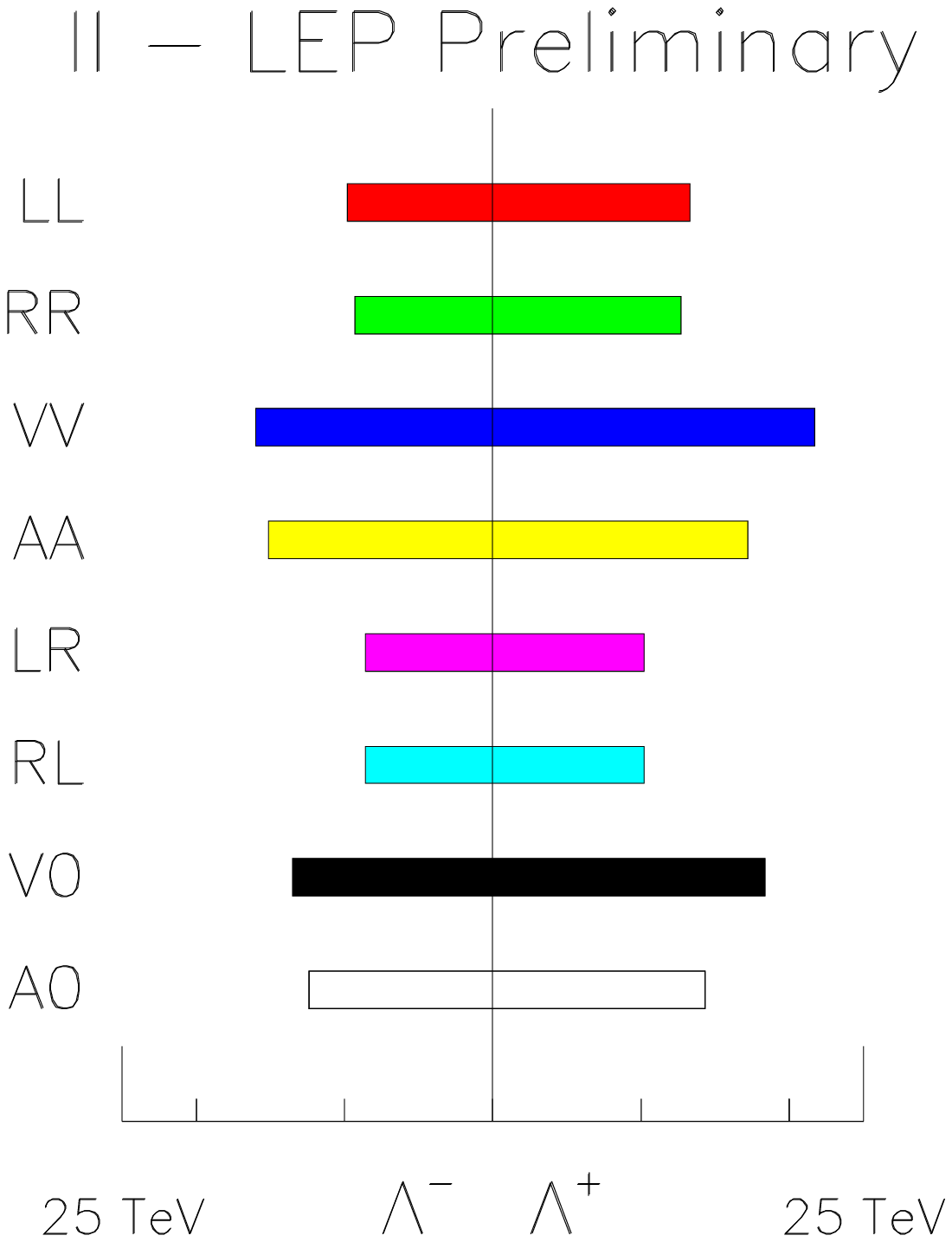}
\includegraphics{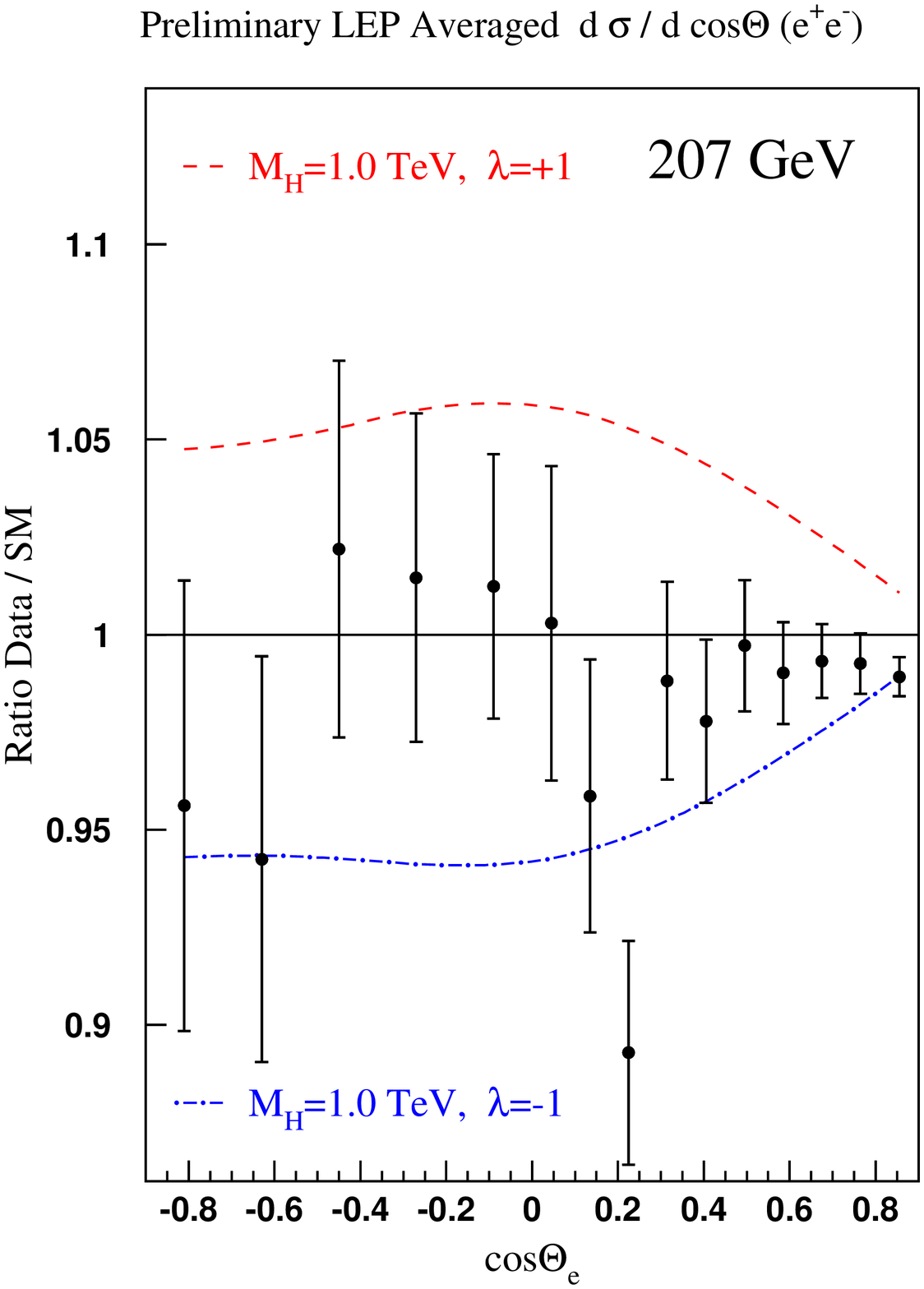}
 \caption{\it Left: Examples for limits on contact interactions for $\mu$ 
plus $\tau$ final states. 
Right: Comparison of the ${\mathrm e^+e^-}$ final state differential cross-section with the 
Standard Model expectation and with predictions 
including virtual graviton exchanges at a mass scale of 1~TeV.  
    \label{gravity} }
\end{figure}

In the framework of contact interactions it is assumed that LEP2 has sensitivity to additional 
contributions to the Langragian which are parametrized in the form
\begin{equation}
{\mathcal L}_{\mathrm eff} = \frac{g^2}{(1+\delta)\Lambda^2} \sum_{i,j=L,R}\eta_{ij} 
        {\overline {\mathrm e}}_i \gamma_{\mu} {\mathrm e}_i {\overline {\mathrm f}}_j \gamma^{\mu} {\mathrm f}_j
\end{equation}
with a coupling strength $g$ for which usually $g^2/4\pi = 1$ is assumed, with $\Lambda$
as the mass scale of the interaction, and with $\delta =$~1~(0) 
for ${\mathrm f = e}$ ($\mathrm f \neq e$). 
Also contained in the formula are the possibilities
to define the helicity of the currents ($L$, $R$) and the 
sign of the interference of the new physics 
contributions with the Standard Model processes, $\eta_{ij} = \pm$1. For the various models that can 
be built from this Langrangian, the predictions are fitted to the measured total and/or 
differential cross-sections and asymmmetries; the fitting parameter is defined as 
$\epsilon := 1/\Lambda^2$. The fitted $\epsilon$ values are converted into 95$\%$ 
confidence level limits on $\Lambda$ which for all models are larger than 2.1 to 21.7~TeV, 
depending on the model. It is therefore fair to say that no signs for contact interactions have 
been found at LEP2. The left part of figure~\ref{gravity} shows an example for the limits 
in the $\mu$ plus $\tau$ channel for various models.

${\mathrm Z'}$ bosons are an ingredient of several new physics models: In the E6 Grand 
Unified Theory, the group structure breaks down to the known Standard Model 
$SU(3)_c \times SU(2)_L \times U(1)_Y$, but additional U(1) subgroups are 
also present which lead to ${\mathrm Z'}$ bosons. The Sequential Standard Model postulates 
the existence of a ${\mathrm Z'}$ boson with the same couplings as the 
standard ${\mathrm Z}$ boson\footnote{The mixing between the ${\mathrm Z}$ and the ${\mathrm Z'}$ is 
found to be consistent with 0, as is expected from LEP1 where no sign of ${\mathrm Z'}$ was 
found in the precision measurements of the ${\mathrm Z}$.}. 
And in the so-called Left-Right Model, the existence of an additional 
$SU(2)_R$ subgroup would lead to ${\mathrm Z'}$ and ${\mathrm W'^{\pm}}$ bosons. 
The effect of the ${\mathrm Z'}$ can be parametrised as a contact term in the cross-section, so that 
the modified cross-section prediction can be fitted to the data, aiming for the extraction of
the mass $M_{{\mathrm Z'}}$ of the new boson. Depending on the model considered the derived limit 
for ${\mathrm Z'}$ bosons is found to be between 434 and 1787~GeV and therefore beyond the reach of LEP2. 

Quantum gravity is a candidate for solving the hierarchy problem, i.e. for providing the 
missing link between the electroweak scale of order ${\mathcal O}$(1~TeV) and the Planck scale
$M_{Pl} = 10^{16}$~TeV. Assuming that quantum gravity lives in 4+n dimensions, whereas our Standard
Model phenomenology is confined to the usual 3+1 dimensions, it is possible that the 4+n 
quantum gravity pendant to the Planck scale is of the order of the electroweak scale. 
This new `Planck scale' $M_D$ would be related to the usual Planck scale $M_{Pl}$ via
\begin{equation}
M_{Pl}^2 = M_D^{2+n} \cdot R^n
\end{equation}
with  $R$ the size of the new additional dimensions. 
The effect of quantum gravity would be the exchange of virtual gravitons, a process which would
interfere with the Standard Model ${\mathrm Z^0}$ exchange and would thus modify the cross-sections 
measured at LEP2. The cross-section including the effect of graviton exchange is of the form
\begin{equation}
\frac{d\sigma}{d\cos\theta} = SM + A \cdot \frac{\lambda}{M_H^4} + B \cdot (\frac{\lambda}{M_H^4})^2 
\end{equation} 
with a Standard Model term, an interference term and a pure new physics term with an amplitude 
proportional to $\lambda/M_H^4$ (for contact interactions the new physics amplitude is
proportional to the inverse of the mass scale squared). The parameter $\lambda$, which 
cannot be known without the knowledge of the full quantum gravity theory, is usually set to $\pm$1.
Then a fit depending on the parameter
$\epsilon = \lambda / M_H^4$ is performed, and a limit on the 
mass scale $M_H$, which is related to the 
new Planck scale $M_D$, can be derived. At the 95$\%$ 
confidence level, $M_H >$~1.20~TeV (1.09~TeV) for 
$\lambda =$~+1 (-1). So also signs of quantum gravity cannot be seen at LEP2. 
Figure~\ref{gravity}, right, shows the ratio of the measured differential 
${\mathrm e^+e^-} \rightarrow {\mathrm e^+e^-}$ cross-section
at 207~GeV together with the Standard Model expectation. 
The data are well compatible with 1. The two model 
predictions for the virtual graviton exchange with a graviton mass of 1~TeV 
and $\lambda = \pm$1, however, clearly fail to describe the data. 

\section{\label{conclusion}Conclusions and Outlook}
The LEP and SLD heavy flavour electroweak measurements have become very stable over the 
past few years and are almost all finalized. The interpretation in the electroweak Standard 
Model leads to an interesting discrepancy  of 2.9$\sigma$
between hadronic and leptonic measurements of the 
effective electroweak mixing angle for leptons, 
$\sin^2\theta_{\mathrm eff}^{\mathrm lept}$, which 
is basically due to a discrepancy between the SLD ${\mathcal A}_{\mathrm l}$ and 
the LEP $A_{{\mathrm FB}}^{\mathrm b}$ contributions to 
the average. To discover the origin of this discrepancy will be a task for future colliders. 

The measurements of 2-fermion processes at LEP2 energies are in a good overall agreement with 
the Standard Model expectation. The cross-sections and asymmetries measured by the LEP 
collaborations are nevertheless used to extract limits on new physics beyond the Standard Model. 
Models considered include contact interactions, 
${\mathrm Z'}$ bosons, and low scale gravity. No signs of
new physics at LEP are observed. The limits for new physics are mostly well above 
1~TeV.  
\section{Acknowledgements}
I would like to thank the LEP/SLD heavy flavour and 2-fermion 
electroweak working groups for the results presented in this review. 
In addition, I would like to thank P.~Wells, K.~Sachs, 
R.~Hawkings and M.~Elsing for 
their critical reading of this text, and the La Thuile 
conference organizers for their hospitality. 

\begin{thebibliography}{9}
\bibitem{rblit}
M.~Acciarri {\it et al.}, Eur. Phys. J. {\bf C13} 47 (2000);
G.~Abbiendi {\it et al.}, Eur. Phys. J. {\bf C8} 217 (1999);
K.~Abe {\it et al.}, Phys. Rev. Lett. {\bf 80} 660 (1998);
R.~Barate {\it et al.}, Eur. Phys. J. {\bf C4} 557 (1998);
R.~Barate {\it et al.}, Eur. Phys. J. {\bf C16} 597 (2000);
P.~Abreu {\it et al.}, Eur. Phys. J. {\bf C12} 209 (2000);
P.~Abreu {\it et al.}, Eur. Phys. J. {\bf C12} 225 (2000);
K.~Ackerstaff {\it et al.}, Eur. Phys. J. {\bf C1} 439 (1998);
G.~Alexander {\it et al.}, Z. Phys. {\bf C72} 1 (1996);
SLD Collaboration, paper 174 contributed to ICHEP98 Vancouver, Canada.   
%
\bibitem{opalafbb}
G.~Abbiendi {\it et al.}, Phys. Lett. {\bf B546} 29 (2002).
%
\bibitem{delphiafbl}
DELPHI Collaboration, DELPHI-2002-028 CONF 562.
%
\bibitem{afblit}
M.~Acciarri {\it et al.}, Phys. Lett. {\bf B439} 225 (1998);
M.~Acciarri {\it et al.}, Phys. Lett. {\bf B448} 152 (1999); 
DELPHI Collaboration, DELPHI-2001-048 CONF 476;
A.~Heister {\it et al.}, Eur. Phys. J {\bf C22} 201 (2001).
%
\bibitem{afblit2}
P.~Abreu {\it et al.}, Eur. Phys. J. {\bf C10} 219 (1999);
G.~Alexander {\it et al.}, Z. Phys. {\bf C73} 379 (1996);
A.~Heister {\it et al.}, Eur. Phys. J {\bf C24} 177 (2002); 
OPAL Collaboration, OPAL Physics Note 226.
%
\bibitem{afbc}
R.~Barate {\it et al.}, Phys. Lett. {\bf B}434 (1998) 415.
%
\bibitem{2flit1}
LEP Electroweak Working Group ${\mathrm f}\overline{{\mathrm f}}$ Subgroup, LEP2FF/02-03.
%
\bibitem{2fweb}
http://www.cern.ch/LEPEWWG/lep2/
%
\end{thebibliography}
\end{document}